\documentclass[12pt]{elsart}
\usepackage{graphicx}
\usepackage{amsmath}
\begin{document}
\runauthor{Eduard P. Kontar}
\begin{frontmatter}
\title{Numerical consideration of quasilinear electron cloud dynamics in plasma}
\author[]{Eduard P. Kontar}
\address{Institute of Theoretical Astrophysics, P.O. Box 1029, Blindern,
0315 Oslo, Norway, E-mail: {\tt eduardk@astro.uio.no}}
\begin{abstract}
The dynamics of a hot electron cloud in the solar corona-like
plasma based on the numerical solution of kinetic equations of
weak turbulence theory is considered. Different finite difference
schemes are examined to fit the exact analytical solutions of
quasilinear equations in hydrodynamic limit (gas-dynamic
solution). It is shown that the scheme suggested demonstrates
correct asymptotic behavior and can be employed to solve initial
value problems for an arbitrary initial electron distribution
function.

\end{abstract}
\begin{keyword}
Sun; plasma; electron beams; finite difference method;
\PACS{95.30.Q; 52.35; 41.75.F; 02.70.B}
\end{keyword}
\end{frontmatter}

\section{Introduction}

 Accelerated particle beams occur in the wide range of astrophysical
 situations as solar flares, cosmic rays, radio jets,
 magnetospheres of pulsars, planetary atmospheres, etc \cite{Kaplan,Benz}. The bright
 signatures of the electron beams in a plasma are the solar type
 III bursts \cite{Suzuki,Dulk}.  In accordance with current understanding of these
 bursts an electron beam
 propagating along open magnetic field lines from the Sun toward the Earth
 generate Langmuir waves, which are partly transformed into observable radio
 emission via nonlinear plasma processes \cite{Suzuki,Melrose1990}.
 The typical density of electron beams is low and
 Coulomb collisions have no influence on beam dynamics. The main process of
 beam interaction with the surrounding plasma is resonant Cherenkov's generation
 and absorption of plasma waves  \cite{Melrose1990}.
 The Langmuir waves excited by the electron beam flatter the
 electron distribution function \cite{Vedenov1970}.
 Thus, for the characteristic time of electron-wave interaction (quasilinear time)
 $\tau \approx n'/\omega _{pe} n$ (where $n'$, $n$ are the beam an plasma density,
 and $\omega _{pe}=\sqrt{e^2n/\epsilon _0m}$ is the electron plasma
 frequency) plateau is formed at the electron distribution function
\cite{Vedenov1970}. Propagation of electrons disturbs local
equilibrium  and in the next spatial point generation of waves
repeats. Many authors considered the problem of electron beam
propagation analytically
\cite{Ryutov,Zaitsev,Grognard1975,Shibahashi,Melnik} as well as
numerically
\cite{Smith,Takakura,Grognard,McClements,Muschietti,KLM}. However,
the results obtained are far from quantitative agreement
\cite{Melrose1990}. This is mainly connected with the fact that
the system of kinetic equations describing the problem is
nonlinear with stiff relaxation terms. The problem can be
significantly simplified if the smallness of quasilinear time can
be taken into account as it was suggested by Ryutov and Sagdeev
\cite{Ryutov}. Thus, implying that plateau is established at the
electron distribution function and high level of plasma waves is
generated at every spatial point one can turn from kinetic to
gas-dynamic description. The gas-dynamic system of equation
describing electron cloud dynamics in hydrodynamic limit has been
recently derived by Mel'nik \cite{Melnik}. The solution obtained
in \cite{Melnik,MLK} for the initial distribution function
$\partial g_0(v)/\partial v>0$ is a compact object propagating in
a plasma with conservation of the particle number, energy, and
wave energy. However, the solutions obtained were not supported by
numerical calculations. It is argued in \cite{Grognard1980} that
it is even impossible to obtain the stable solution for initially
unstable electron beam. On the contrary, gas-dynamic description
\cite{Melnik,MLK} demonstrates that initially unstable electron
beam can lead to interesting solutions.

Numerical consideration of quasilinear equations with initial
distribution function $\partial g_0(v)/\partial v <0$ has been
conducted several times
\cite{Shibahashi,Smith,Takakura,Grognard,McClements,MLK}.

In the given paper the numerical solution of kinetic equation of
quasilinear theory is discussed. The dynamics of the electron beam
is considered at the scale much larger that the size of electron
beam. We show that the transport term should be approximated by
the higher than first order finite difference operator. The
different presentation of collision terms are examined. The
results of numerical solution are presented for typical parameters
of electron beams and solar plasma are presented.

\section{The statement of the problem}
Let us consider the propagation of the electron beam cloud when
the energy density of excited Langmuir waves is much less than
that of surrounding plasma
\begin{equation}
W/nT \ll (k\lambda _{D})^2,
 \label{eq1}
\end{equation}
where $W$ is the energy density of Langmuir waves, $T$ is the
temperature of surrounding plasma, $k$ is the wave number, and
$\lambda _{D}=\sqrt{k_BT\epsilon _0/ne^2}$ is electron Debye
length. Our analysis is limited by one-dimensional kinetic
equations following \cite{Ryutov,Zaitsev}. One-dimensional beam
propagation is supported by numerical solution of 3D equations
\cite{Agapov}. In the application to solar burst III type those
electrons propagate along magnetic field which energy $\mu _0H^2/2
\gg nmv_0^2/2$ \cite{Suzuki}, that ensures one-dimensional
character of electron propagation.

In the case of type III bursts, as it was shown by Vedenov,
Velikhov and Sagdeev \cite{Vedenov} and by Drummond and Pines
\cite{Drumond}, one can use equations of quasilinear theory
\cite{Ryutov}
\begin{equation}
\frac{\partial f}{\partial t}+v \frac{\partial f}{\partial x}=
\frac{e^2 }{\epsilon _0 m^2}\frac{\partial}{\partial v}
\frac{W}{v}\frac{\partial f}{\partial v}, \label{eq:2a}
\end{equation}
\begin{equation}
\frac{\partial W}{\partial t}=\frac{\pi
\omega_{pe}}{n}v^2W\frac{\partial f}{\partial v}, \;\;\;
\omega_{pe}=kv \label{eq:3a}
\end{equation}
where $f(v,x,t)$ is the electron distribution function, $W(v,x,t)$
is the spectral energy density of Langmuir waves. $W(v,x,t)$ plays
the same role for waves as the electron distribution function does
for particles. The system (\ref{eq:2a},\ref{eq:3a}) describes the
resonant interaction $\omega_{pe}=kv$ of electrons and Langmuir
waves, i.e. electron with the velocity $v$ can emit or absorb a
Langmuir wave with the phase velocity $v_{ph}=v$.  The group
velocity of Langmuir waves is small as $v_g\approx v_{Te}^2/v \ll
v$ and therefore the corresponding term in the left side of the
equation (\ref{eq:3a}) is omitted \cite{Ryutov}.

The clouds of fast electrons are formed in the spatially limited
regions of solar corona where acceleration takes place. Therefore,
spatially bounded beam is taken for consideration. The initial
electron distribution function is
\begin{equation}
F(v,x,t=0)=g_0(v)\mbox{exp}(-x^2/d^2), \label{eq:4}
\end{equation}
where $d$ is the characteristic size of the electron cloud and
$g_0(v)$ is the initial distribution of electrons in the velocity
space and $\int\limits_0^{\infty}g_0(v)dv =n'$. It is also implied
that initially the spectral energy density of Langmuir waves is of
the thermal level and homogeneously distributed in space
\begin{equation}
W(v,x,t=0)=10^{-8}mn'v_0^3/\omega_{pe}, \label{eq:5}
\end{equation}
where $v_0$ is some characteristic velocity of the electron cloud.
The system of kinetic equations (\ref{eq:2a},\ref{eq:3a}) is
nonlinear with two characteristic time scales. The first is the
quasilinear time $n/n'\omega_{pe}$ that is determined by the
interaction of particles and waves. The second scale is the time
length of the electron cloud $d/v_0 >>n/n'\omega_{pe}$. And we are
interested in dynamics of the electron cloud at the time scale $t
>>d/v_0$.

\section{Numerical method}

The problem we are confronted with is an initial value problem.
There are a variety of techniques available for the numerical
solution of such partial differential equations. We use finite
differencing (see, for example, \cite{SmithGD,Samarskii,Thomas}).
For our further consideration we rewrite partial differential
equations together with initial conditions in the following form
\begin{equation}
\frac{\partial F}{\partial t}+ \alpha V\frac{\partial F}{\partial
X}= \frac{1}{\tau}\frac{\partial}{\partial V} D\frac{\partial
F}{\partial V}, \label{eq:2d}
\end{equation}
\begin{equation}
\frac{\partial D}{\partial t}=\frac{1}{\tau}V^2D\frac{\partial
F}{\partial V}, \label{eq:3d}
\end{equation}
\begin{equation}
F(V,X,t=0)=G_0(V)\mbox{exp}(-X^2),\;\;
\mbox{where}\;\;\int^{\infty}_0G_0(V)dV =1, \label{eq:4d}
\end{equation}
\begin{equation}
W(V,X,t=0)=10^{-8}, \label{eq:5d}
\end{equation}
where we use normalized velocity $V=v/v_0$, distance $X=x/d$,
quasilinear time $\tau =n/n'\omega_{pe}\pi$, electron distribution
function $F(V,X,t)=f(v,x,t)v_0/n'$, and
$D(V,X,t)=W(v,x,t)\omega_{pe}/vmn'v_0^3$, $\alpha =v_0/d$. All
terms in equations (\ref{eq:2d},\ref{eq:3d}) are presented in 1/s
units.

For numerical solution of the equation (\ref{eq:2d},\ref{eq:3d})
we will introduce the grid points $V_j$, for $j=1,...,M$, and
$X_i$ for $i=1,...,N$ with uniform mesh width $\Delta V
=V_{j+1}-V_j$, $\Delta X =X_{i+1}-X_{i}$. The discrete time level
$t_k$ is also uniformly spaced with the time step $\Delta t$.
Given any function $U$ we denote its nodal value by
$U_{j+1/2}=U(y_{j+1/2})$ and its cell average values by
\begin{equation}\label{eq:cell}
  U_j =\frac 1{\Delta y}\int_{y_{j-1/2}}^{y_{j+1/2}}U(y)dy,
\end{equation}
where $y$ is an variable ($X$ or $V$). We also introduce operators
\begin{equation}\label{eq:oper}
  \nabla ^{+}_jU\equiv
  \frac{U_{j+1}-U_{j}}{y_{j+1}-y_{j}}, \;\;\;\;\;\;\;
   \nabla ^{-}_jU\equiv \frac{U_{j}-U_{j-1}}{y_{j}-y_{j-1}}.
\end{equation}
The use of operator splitting \cite{Hamilton} allows us to solve
equations (\ref{eq:2d}, \ref{eq:3d}) by determining the finite
difference operators for each individual term of the equation. We
now describe the finite difference schemes used to advance each
term.

\subsection{Quasilinear relaxation}

 It case the electron beam is homogeneously distributed in space
 transport term in (\ref{eq:2d}) should be omitted. Thus we obtain the
 system of equations that describes quasilinear relaxation of
 an electron beam in velocity space
\begin{equation}
\frac{\partial F}{\partial t}=
\frac{1}{\tau}\frac{\partial}{\partial V }D \frac{\partial
F}{\partial V}, \label{eq:2q}
\end{equation}
\begin{equation}
\frac{\partial D}{\partial t}=\frac{1}{\tau}V^2D \frac{\partial
F}{\partial V}, \label{eq:3q}
\end{equation}
The equations(\ref{eq:2q},\ref{eq:3q}) can be solved analytically
\cite{Vedenov1970}. Thus substituting (\ref{eq:3q}) into
(\ref{eq:2q}) one obtains quasilinear integral
\begin{equation}\label{eq:int}
  \frac{\partial}{\partial t}\left[F - \frac 1{V^2}\frac{\partial
  D}{\partial V}\right] =0,
\end{equation}
that allows us to obtain final distribution of particles and waves
via initial conditions. The initially unstable electron
distribution function $\partial G_0(V)/\partial V>0$ leads to
generation of plasma waves and flattering of the electron
distribution function. Quasilinear relaxation continues until
$\partial F(V)/\partial V=0$, and plateau is formed at the
electron distribution function. The characteristic time of
beam-plasma interaction is $\tau$ (for this time a half of initial
electron beam energy is transferred into waves).

For initially monoenergetic electron beam $G_0(V)=\delta (V-1)$
using (\ref{eq:int}) \cite{Vedenov1970} one finds the following
steady state solution (the solution of the system at $t\rightarrow
\infty $)
\begin{equation}\label{eq:sol1}
  F_{\infty}(V)=\left\{
  \begin{array}{ll}
              1,&\mbox{$V<1$}\\
              0,&\mbox{$V>1$}
              \end{array}
\right.\;\;\;\;\;
 D_{\infty}(V)=\left\{
  \begin{array}{ll}
              V^3,&\mbox{$V<1$}\\
              0,&\mbox{$V>1$}
              \end{array}
\right.
\end{equation}
Similar to (\ref{eq:sol1}) for the initial distribution function
$G_0(V)=2V$, for $V<1$ which has been considered in \cite{MLK} one
obtains the following steady state solution
\begin{equation}\label{eq:sol2}
  F_{\infty}(V)=\left\{
  \begin{array}{ll}
              1,&\mbox{$V<1$}\\
              0,&\mbox{$V>1$}
              \end{array}
\right.\;\;\;\;\;
 D_{\infty}(V)=\left\{
  \begin{array}{ll}
              V^3(1-V),&\mbox{$V<1$}\\
              0,&\mbox{$V>1$}
              \end{array}
\right.
\end{equation}
The solution of quasilinear equations (\ref{eq:2q}, \ref{eq:3q})
shows that the maximum of the spectral energy density depends on
the initial electron distribution function. The maximum of $D(V)$,
$D=1$ is reached at $V=1$ when the initial electron distribution
function is an monoenergetic beam.

 To construct an conservative finite difference scheme we follow
 \cite{Samarskii}. For equation (\ref{eq:2q}) one writes the equation
 of balance in the cell
 $V_{j-1/2}\leq V\leq V_{j+1/2}$, and $t_k\leq t \leq t_{k+1}$
\begin{equation}\label{eq:bal}
  \int_{V_{j-1/2}}^{V_{j+1/2}}\left[F(V,t_{k+1})-F(V,t_k)\right]dV
  =
  \int_{t_{k}}^{t_{k+1}}\left[w(V_{j+1/2},t)-w(V_{j-1/2},t)\right]dt,
\end{equation}
where
\begin{equation}\label{eq:w0}
  w(V,t)=\frac{D}{\tau}\frac{\partial F}{\partial V},
\end{equation}
is the particle flux in velocity space. Integrating each term in
(\ref{eq:bal}) and using that
\begin{equation}\label{eq:w2}
  w_{j+1/2}=a_j \nabla^{+}_{j}F_j,
\end{equation}
\begin{equation}\label{eq:w1}
 \frac 1{\Delta t}\int_{t_{k}}^{t_{k+1}}w(V_{j+1/2},t)dt =\sigma
 w_{j}^{k+1}+(1-\sigma)w_{j}^{k},
\end{equation}
one obtains the following general finite-difference equation
\begin{equation}\label{eq:g}
  F_j^{k+1}-F_{j}^{k}=
  \nabla_{j}^{-}a_j\nabla_{j}^{+}\left(\sigma F_{j}^{k+1}+(1-\sigma)F_{j}^k
  \right),
\end{equation}
where $\sigma$ is a number $0\leq \sigma \leq 1$,and $a_j(D)$ is a
functional
\begin{equation}\label{eq:a}
a_j = \left(\frac{\tau}{\Delta
V}\int_{V_{j}}^{V_{j+1}}\frac{dV}{D(V)}\right)^{-1}=\left[\tau\int_{0}^{1}\frac{ds}{D(V+\Delta
Vs )}\right]^{-1},
\end{equation}
which can be approximated in a number of different ways
\cite{Samarskii}.

Let us consider a few interesting cases. For $\sigma =1$ and
$a_j=D_j/\tau$ we rederive the fully implicit scheme used by
Grognrad \cite{Grognard}
\begin{equation}\label{eq:gr1}
  F_j^{k+1}-F_{j}^{k}=\frac{\Delta t}{\tau}
  \nabla_{j}^{-}D_j^{k}\nabla_{j}^{+}F_{j}^{k+1}
\end{equation}
\begin{equation}\label{eq:gr2}
  D_j^{k+1}-D_{j}^{k}=\frac{\Delta t}{\tau}V_j^2 D_j^k\nabla_{j}^{+}F_{j}^{k+1}
\end{equation}
Equation (\ref{eq:gr1}) is unconditionally stable whereas second
one (\ref{eq:gr2}) is stable only when $\Delta t\leq \tau/(2V_j^2
\nabla_{j}^{+}F_{j}^{k+1})$. Using (\ref{eq:gr1}) and
(\ref{eq:gr2}) together one obtains unbounded growth of the
Langmuir waves \cite{Grognard}.  To solve the problem
 spontaneous terms have been added to the right
hand side of equations (\ref{eq:gr1}) and (\ref{eq:gr2})
\cite{Grognard}.

The scheme (\ref{eq:gr1},\ref{eq:gr2}) is not acceptable when
spontaneous terms are omitted \cite{Grognard} (as in our case).
Indeed, in absence of spontaneous terms we know that the
relaxation is non-linear diffusion in velocity space which
decreases $|\partial F/\partial V|$ in a finite domain but
increases $|\partial F/\partial V|$ to very large values at the
limits of this domain.

For $\sigma =0$ one obtains fully explicit schemes. If the
functional is approximated as $a_j=D_j/\tau$ we obtain the scheme
is used by Takakura \cite{Takakura}(hereafter scheme I)
\begin{equation}\label{eq:Ia}
  F_j^{k+1}-F_{j}^{k}=\frac{\Delta t}{\tau}
  \nabla_{j}^{-}D_j^{k}\nabla_{j}^{+}F_{j}^{k}
\end{equation}
\begin{equation}\label{eq:Ib}
  D_j^{k+1}-D_{j}^{k}=\frac{\Delta t}{\tau}V_j^2 D_j^k\nabla_{j}^{+}F_{j}^{k}
\end{equation}
and more accurate scheme (scheme II), when
$a_j=(D_{j+1}+D_j)/2\tau$
\begin{equation}\label{eq:IIa}
  F_j^{k+1}-F_{j}^{k}=\frac{\Delta t}{2\tau}
  \nabla_{j}^{-}(D_{j+1}^{k}+D_{j}^{k})\nabla_{j}^{+}F_{j}^{k}
\end{equation}
\begin{equation}\label{eq:IIb}
  D_j^{k+1}-D_{j}^{k}=\frac{\Delta t}{\tau}V_j^2 D_j^k\nabla_{j}^{-}F_{j}^{k}
\end{equation}
Both schemes are conditionally stable. The criteria of stability
is
\begin{equation}\label{eq:dt}
  \Delta t\leq \mbox{min}\left[ \tau /(2V_j^2\nabla_{j}^{+}F_{j}^{k+1}),\;\;
  \tau \Delta V^2/2a_j(V_j)\right]\;\; \mbox{for}\;\;1\leq j\leq M, \;\; k>0
\end{equation}
Using that $D$ is always less than $1$, and
$\nabla_{j}^{+}F_{j}^{k+1}\leq 1/\Delta V^2$ for an arbitrary
initial distribution function timestep should be as small as
$\Delta t\leq\tau \Delta V^2/2$.

\begin{figure}
\includegraphics[width=130mm]{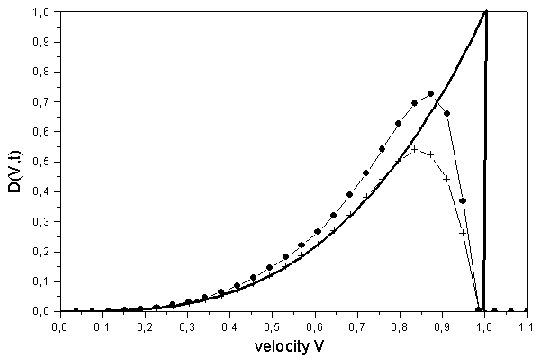}
\includegraphics[width=130mm]{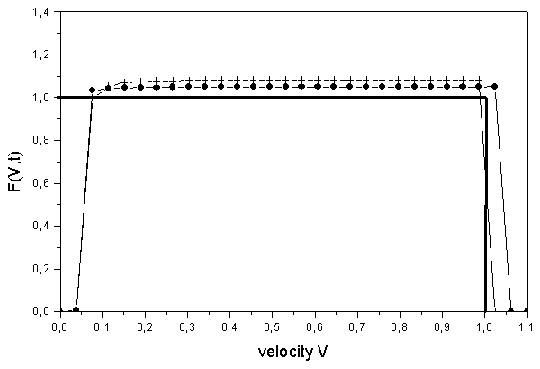}
 \caption{Spectral energy density and electron distribution function at $t=1.0$s. Results
 of numerical solution of the system (\ref{eq:2q},\ref{eq:3q})
 for initial monoenergetic distribution function  $G_0(V)=2\mbox{exp}(-(V-1)^2/\Delta V_0^2)/\sqrt{\pi}\Delta V_0$ for $V\leq
1$ ($\Delta V_0 =1/8$, $\tau =0.02$s, $\Delta V =0.038$, $\Delta
t=1.4\times 10^{-5}$s). Scheme I (plus signs), scheme II (black
circles), and theoretical solution (\ref{eq:sol1}) (solid line)}
 \label{fig1}
\end{figure}

\begin{figure}
\includegraphics[width=130mm]{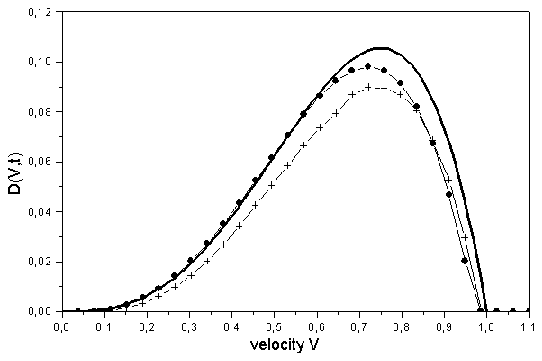}
\includegraphics[width=130mm]{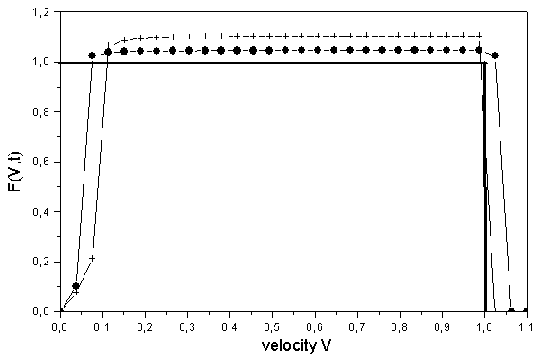}
\caption{Spectral energy density and electron distribution
function at $t=1.0$s. Results
 of numerical solution of the system (\ref{eq:2q},\ref{eq:3q})
 for initial distribution function $G_0(V)=2V$ for $V\leq 1$ ($\tau =0.02$s,
 $\Delta V =0.038$, $\Delta t=1.4\times 10^{-5}$s). Scheme I
 (plus signs), scheme II (black circles), and theoretical
 solution (\ref{eq:sol1}) (solid line)}
  \label{fig2}
\end{figure}

\begin{figure}
\includegraphics[width=130mm]{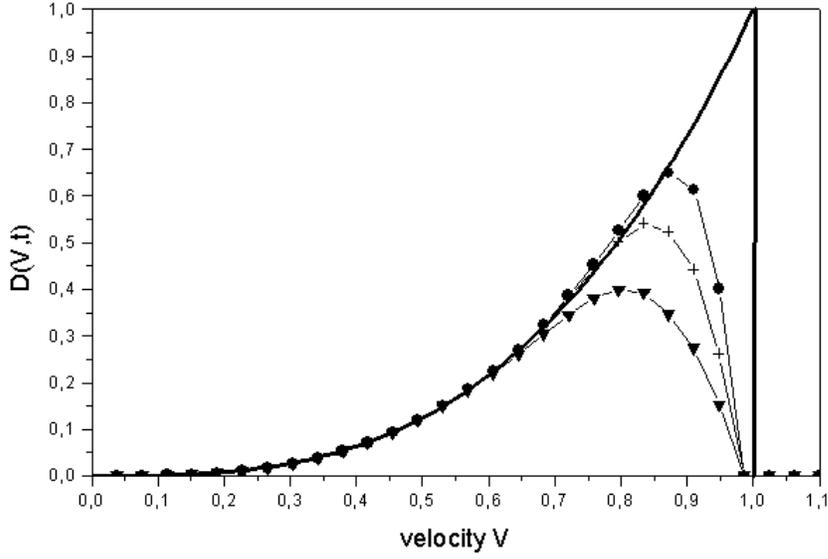}
\caption{Spectral energy density at $t=1.0$s for various electron
beams. Results of numerical solution of the system
(\ref{eq:2q},\ref{eq:3q})
 for initial distribution function $G_0(V)=2\mbox{exp}(-(V-1)^2/\Delta V_0^2)/\sqrt{\pi}\Delta V_0$ for $V\leq
1$ ($\tau =0.02$s, $\Delta V =0.038$, $\Delta t=1.4\times
10^{-5}$s).  $\Delta V_0=1/5$ (triangle signs), $\Delta V_0=1/8$
(plus signs), $\Delta V_0=1/12$ (black circles) and theoretical
curve (\ref{eq:sol1}) (solid lines).}
  \label{fig3}
\end{figure}

   To test the schemes I and II we use typical beam-plasma parameters
that ensures smallness of $\tau$. The results of numerical tests
are presented in fig. \ref{fig1}-\ref{fig2}. Test run shows that
both schemes correctly approximate the process of quasilinear
relaxation. The scheme II gives better approximation for both
initial electron distribution functions presented in this section.
In fig. \ref{fig2} we obtain the best coincidence between
numerical results and the analytical solution (\ref{eq:sol2}). In
case of initially monoenergetic electron beam (actually
$G_0(V)=2\mbox{exp}(-(V-1)^2/\Delta V_0^2)/\sqrt{\pi}\Delta V_0$
for $V\leq 1$, $\Delta V_0\ll 1$) we have the correct asymptotic
behavior. Decreasing initial dispersion in the beam $\Delta V_0 $
we obtain the spectral energy density approaching to the
analytical solution (\ref{eq:3q}) (see \ref{fig3}). Scheme I gives
higher plateau but more accurate drop of distribution function at
$V=1$, whereas scheme II better approximates plateau but has more
smooth border at $V=1$. From fig. \ref{fig1}, \ref{fig2} we can
see that scheme II lead to appearance of "accelerated" particles
($F$ is different from zero at $V=1+\Delta V$). Consequently, if
we take into account transport of particles the maximum velocity
of the plateau will unphysically grow at beam propagation. Thus,
scheme I may be applied for short-time beam dynamics at the time
scale $t\leq d/v_0$ whereas scheme II is more suitable for long
time dynamics $t\gg d/v_0$.


\subsection{Transport of particles}

Propagation term plays an important role for initial distribution
function $\partial G_0(V)/\partial V <0$. The equation to consider
is
\begin{equation}\label{eq:free}
\frac{\partial F}{\partial t}+ \gamma\frac{\partial F}{\partial
X}=0,\;\;\; \gamma =\mbox{const}>0
\end{equation}
Since dynamic calculations at $t>>\tau$ consume much computer time
and therefore finite difference schemes for
(\ref{eq:2d},\ref{eq:3d}) are usually taken as simple as possible.
First order upwind representation of transport operator
(\ref{eq:free}) is taken in the majority of cases
\cite{Shibahashi,Smith,Takakura,Grognard,McClements}
\begin{equation}\label{eq:upwind}
  F_i^{k+1}-F_i^{k}=\beta \left(F_j^k-F_{i-1}^k\right), \;\;\; \beta=\gamma\frac{\Delta t}{\Delta X}
\end{equation}
 However, this first order scheme seems to be not enough to caught
the correct asymptotic behavior of the system at $t\gg d/v_0$. It
is  shown \cite{Leer,Leer1,Leer2} that monotonic transport is the
best finite difference method for equation (\ref{eq:free}). Using
this method one finds that
\begin{equation}
\label{eq:monot1} F_{i}^{k+1}=\left\{
\begin{array}{ll}
F_{i}^{k}-\beta (F_i^k-F_{i-1}^k)-\beta (1-\beta)(\Delta
F_i^k-\Delta F _{i-1}^{k}),\;\;\; \beta>0 \\
F_{i}^{k}-\beta(F_{i+1}^k-F_{i}^k)+\beta (1+\beta)(\Delta
F_{i+1}^k-\Delta F _{i}^{k}).\;\;\; \beta <0
\end{array}
\right.
\end{equation}
where
\begin{equation}
\label{eq:monot2} \Delta F _{i}^{k}=\left\{
\begin{array}{ll}
\displaystyle
\frac{(F_{i}^{k}-F_{i-1}^k)(F_{i+1}^k-F_i^{k})}{(F_{i+1}^{k}-F_{i-1}^{k})},\;\;\;
(F_{i}^{k}-F_{i-1}^k)(F_{i+1}^i-F_i^{k})>0 \\
0,\;\;\;\;\;\;\;\;\;\; \mbox{otherwise}
\end{array}
\right.
\end{equation}
where $F_i^k$ is the value of $F$ at position $X_i$ and the time
$t=k\Delta t$.

In fact we are interested in the evolution of the initial
distribution. Therefore, a good test of the numerical solution is
the propagation of spatially finite distribution. We take an
initial distribution $F(X,t=0)=\mbox{exp}(-X^2)$. In fig.
\ref{fig2} the numerical solutions are compared to the exact
analytical solution of (\ref{eq:free}) (which is just a Gaussian
moving with velocity $\beta$) for upwind (\ref{eq:upwind}) and
monotonic transport (\ref{eq:monot1}) methods. For all methods,
the numerical solution conserves particle number, so that in
general, the height of the numerical solution is a good measure of
accuracy.
\begin{figure}
\includegraphics[width=130mm]{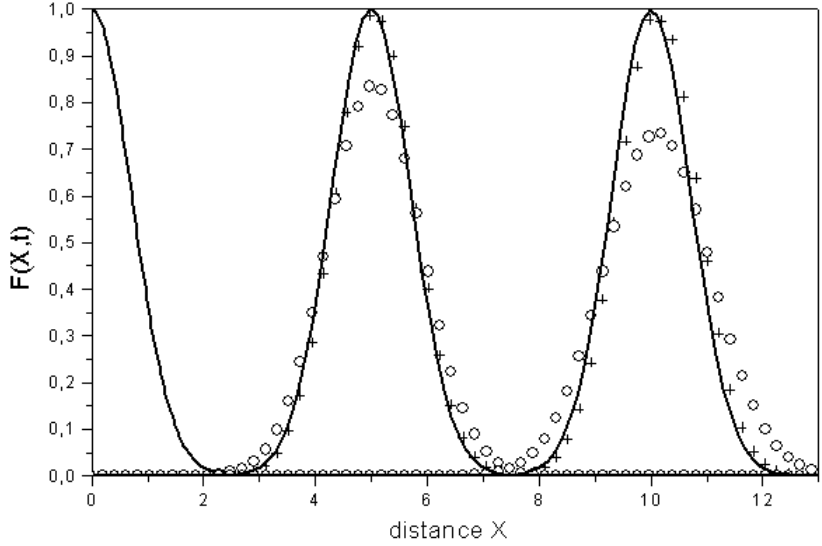}
 \caption{Numerical solution of equation (\ref{eq:free}) with
 initial Gaussian distribution $F(X,t=0)=\mbox{exp}(-X^2)$ at
 $t=1$s, $t=2$s
 ($\Delta X =0.2$, $\gamma =5$ $\Delta t= 1\times10^{-4}$s). Upwind scheme (\ref{eq:upwind})
 (hollow circles) and monotonic scheme (\ref{eq:monot1}) (plus signs).}
 \label{fig4}
\end{figure}

The fact that upwind scheme has "worser" approximation can
drastically influence the electron beam dynamics. Although, the
quasilinear terms are large enough in comparison with the
propagation term, they are of the same order for some points of
$(X,V)$ plain \cite{KLM2000}. In these areas of $(X,V)$ plain the
error of the first order upwind operator can be comparable to the
quasilinear diffusion in the velocity space
\begin{equation}\label{eq:error}
\gamma\frac{ \Delta X}2 \frac{\partial ^2 F}{\partial
  X^2} \approx \frac 1{\tau}\frac {\partial}{\partial
  V}D\frac{\partial F}{\partial V}
\end{equation}
when $D\approx 0$ or $\partial F/\partial V \leq 0$.

On our opinion this can lead to a wrong asymptotic behavior of
numerical solution. Thus, when the initial electron distribution
function is stable as to generation of plasma waves $\partial
G_0(V)/\partial V<0$ the transport term of kinetic equations plays
the main role. The electron propagation causes the change of
electron distribution function and finally the unstable electron
distribution $\partial F/\partial V>0$ appears. The rate of the
quasilinear relaxation strongly depends on beam density at this
point. Therefore, the underestimated electron density will change
the initial point of relaxation \cite{KLM2000}.

\section{Numerical results}

 In this section we consider the long time dynamics of electron
 cloud for typical parameters of the beam and plasma and compare
 to the analytical solution found in hydrodynamic limit (gas-dynamic
 solution \cite{MLK}).

For the time of quasilinear relaxation $\tau$ plateau is formed at
the electron distribution function and the high level of plasma
waves is generated at every spatial point \cite{Vedenov}. The
smallness of quasilinear relaxation time can be used to turn from
kinetic to gas-dynamic description of the problem that was
suggested by Ryutov and Sagdeev \cite{Ryutov} and done by Mel'nik
\cite{Melnik}. Plateau at the electron distribution function and
high level of Langmuir turbulence are implied to be formed. The
method is similar to ordinary hydrodynamics, where we integrate
kinetic equations implying that Maxwell's distribution is assumed
at every point. Following \cite{Melnik,MLK} we can obtain
analytical solution when $\tau \ll t$ for $G_0(V)=2V, \;\;\; v\leq
1$:
\begin{equation}\label{eq:p}
  F(V,X,t)=\mbox{exp}(-(X-\gamma t/2)^2)\theta (1-V)
\end{equation}
\begin{equation}\label{eq:w}
  D(V,X,t)V = V^4\left(1-V)\mbox{exp}(-(X-\gamma t/2)^2)\right)\theta _+(1-V)
\end{equation}
where
\begin{equation}\label{eq:theta}
  \theta (V)=\left\{
  \begin{array}{ll}
              1,&\mbox{$V<0$}\\
              1/2,&\mbox{$V=0$}\\
              0,&\mbox{$V>0$}
              \end{array}
\right.\;\;\;\;\;
  \theta _+(V)=\left\{
  \begin{array}{ll}
              1,&\mbox{$V\leq 0$}\\
              0,&\mbox{$V>0$}
              \end{array}
\right.
\end{equation}
(see \cite{Korn} for details.)

 Electrons (\ref{eq:p}) accompanied by Langmuir waves
(\ref{eq:w}) propagate in a plasma as a beam-plasma structure with
the constant velocity $\gamma /2$.
\begin{figure}
\includegraphics[width=130mm]{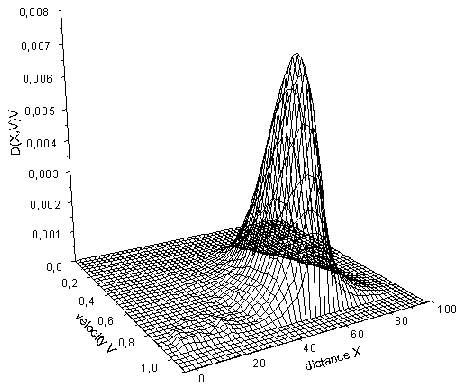}
\includegraphics[width=130mm]{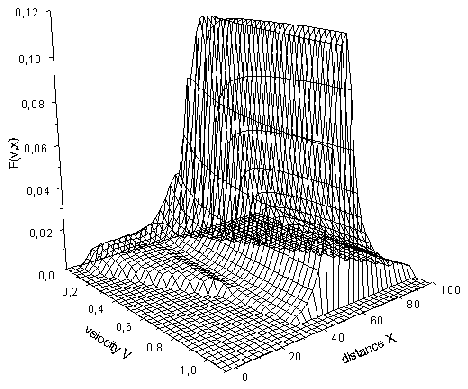}
\caption{Spectral energy density and electron distribution
function at $t=10.0$s. Results
 of numerical solution of the system (\ref{eq:2d},\ref{eq:3d})
 for initial distribution function $G_0(V)=2V$ for $V\leq 1$ ($\gamma =5$,$\tau =0.02$s,
 $\Delta V =0.038$, $\Delta t=1.4\times 10^{-5}$s).}
  \label{fig5}
\end{figure}
\begin{figure}
\includegraphics[width=130mm]{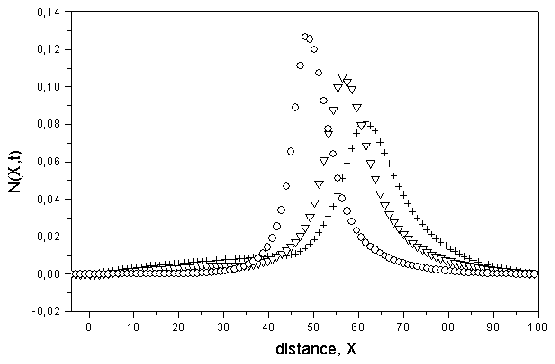}
\caption{Electron beam density at $t=10.0$s for various $\tau$
($\tau =0.0062$s (cross sign), $\tau =0.0025$s (triangle sign),
$\tau =0.0013$s (circle sign)). Results
 of numerical solution of the system (\ref{eq:2d},\ref{eq:3d})
 for initial distribution function $G_0(V)=2V$ for $V\leq 1$, $\Delta V =0.038$,
 $\Delta X =0.2$, $\gamma =5$.}
  \label{fig6}
\end{figure}

The electron distribution function $F(V,X,t)$ and the spectral
energy density of Langmuir waves $D(V,X,t)V$ are presented in fig.
4 at the time moment $t=10.0$s. As it is implied \cite{Melnik,MLK}
plateau is established  in the wide area of velocities from $1$
down to $V\approx 55$ at every spatial point. We notice that the
plateau height exponentially increases on the forward front (which
is presented by the points $X>55$ in fig.~\ref{fig3}) and
exponentially decreases at the back front ($X<55$ ). In the point
( $X\approx 55$ ) the plateau height reaches maximum value.  The
symmetry of the form of the initial electron beam conserves, but
now it is a form of distribution of electron stream (whereas
electrons move with various velocities). In fig.~\ref{fig4} these
Langmuir waves are presented at the moment $t=10$s. Like
electrons, Langmuir waves are concentrated near $X\approx 55$. The
spectral energy density $DV$ reaches its maximum value at
$V\approx 0.8$ that coincides with the theoretical value
(\ref{eq:w}).

 The observable difference between the profile of beam-plasma
structure and the theoretical solution is explained by the fact
that despite the smallness of quasilinear time it is a finite
value $\tau >0$. Some electrons on the tails of the structure do
not take part in quasilinear relaxation and therefore propagate
freely away from the structure. Decreasing the quasilinear time we
can reach better agreement with the analytical solution (fig.
\ref{fig5}).

\section{Conclusions}

The propagation of an electron cloud has complex nonlinear
properties. In order to describe electron propagation in plasma
correctly one needs attention in choose of the method of numerical
solution. The system of partial differential equations has stiff
terms and finite difference scheme should correctly describe
different scales of the system. The different schemes have
different optimal approximations for various time scales. The
analysis of the system shows that first order upwind transport of
the particles may lead to a wrong asymptotic regime. To ensure
sufficient accuracy monotonic scheme has been suggested for
numerical consideration. The monotonic scheme is found to be
accurate enough for the problem considered.

To design schemes with the correct behavior one should use a
discrete analogous of the asymptotic limit of continuous system.
Therefore, the results in asymptotic regimes are compared with
exact analytical solution. Test calculation demonstrate optimistic
agreement for different parameters
 of plasma and a beam. However, the difference between numerical
 solution and gas-dynamic solution is observed. The main assumption of
 the gas-dynamic approach is that plateau is form at every spatial
 point. This is not true for regions of $(X,V)$ plain where the
 fast electron density is low. Quasilinear relaxation for these
 particles is not a fast process and electron propagate in a
 plasma almost freely. The velocities of these electron differ from
 the speed of beam plasma structure and the electrons move away from
 the structure. This leads to the loss of particles and as a
 result the structure becomes lower and wider. However, the mentioned
 difference becomes smaller if we consider the systems with
 smaller $\tau $.

\end{document}